\documentclass[aps,prd,amsmath,amssymb,superscriptaddress,altaffillsymbol, preprintnumbers,nofootinbib,twocolumn]{revtex4}
\pdfoutput=1
\usepackage{amsthm}
\usepackage{graphicx}
\usepackage{color}
\newcommand{\beq}{\begin{eqnarray}}
\newcommand{\eeq}{\end{eqnarray}}

\newcommand{\bmp}{\noindent\begin{minipage}{16cm}}
\newcommand{\emp}{\end{minipage}\vskip 7mm} 

\usepackage{dcolumn}
\usepackage{bm}
\usepackage{bbm}
\usepackage{subfigure}
\usepackage{slashed} 

\usepackage{lipsum}

\theoremstyle{definition}

\theoremstyle{plain}

\usepackage{epsfig}

\usepackage{hyperref}
\definecolor{rossoCP3}{cmyk}{0,.88,.77,.40}
\definecolor{verdeCP3}{rgb}{0.09765625, 0.57421875, 0.1015625}
\definecolor{bluCP3}{rgb}{0, 0.23, 0.67}
\hypersetup{colorlinks, bookmarksopen, bookmarksnumbered,citecolor=bluCP3, linkcolor=bluCP3, pdfstartview=FitH, urlcolor=bluCP3}
\def\lsim{\mathrel{\rlap{\lower4pt\hbox{\hskip1pt$\sim$}}
    \raise1pt\hbox{$<$}}}                
\def\gsim{\mathrel{\rlap{\lower4pt\hbox{\hskip1pt$\sim$}}
    \raise1pt\hbox{$>$}}}                

\newcommand{\bea}{\begin{eqnarray}}
\newcommand{\eea}{\end{eqnarray}}

\newcommand{\ba}{\begin{eqnarray}}
\newcommand{\ea}{\end{eqnarray}}

\newcommand{\be}{\begin{eqnarray}}
\newcommand{\ee}{\end{eqnarray}}

\begin{document}
\title{Self-Interacting Dark Matter through the Higgs Portal}
%
\author{Chris Kouvaris}
\email{kouvaris@cp3-origins.net} 
\author{Ian M. Shoemaker}
\email{shoemaker@cp3-origins.net} 
\affiliation{{\color{black} CP$^{3}$-Origins} \& Danish Institute for Advanced Study {\color{black} DIAS}, University of Southern Denmark, Campusvej 55, DK-5230 Odense M, Denmark}
\author{Kimmo Tuominen}
\email{kimmo.i.tuominen@helsinki.fi}
\affiliation{Department of Physics
and
Helsinki Institute of Physics, P.O.Box 64,
FI-00014, University of Helsinki, Finland}

\begin{abstract}
We study self-interacting dark matter coupled to the Standard Model via the Higgs portal. We consider a scenario where dark matter is a thermal relic with strong enough self interactions that can alleviate the problems of collisionless dark matter. We study constraints from direct detection searches, the LHC, and Big Bang nucleosynthesis. We show that the tension between these constraints and the need for sufficiently strong self-interactions with light mediators can be alleviated by coupling the mediator to either active or sterile neutrinos. {Future direct detection data offers great potential and can be used to find evidence of a light mediator and verify that dark matter scatters via long-range self-interactions.}
\preprint{CP3-Origins-2014-039 DNRF90, DIAS-2014-39, HIP-2014-28/TH.}
 \end{abstract}

\maketitle

\section{Introduction}


There appears to be tension between the observations and simulations of small-scale structure of Collisionless Cold Dark Matter (CCDM).  Observations of dwarf galaxies seem to confirm that the dark matter (DM) density profile is not cuspy as one gets closer to the center of the galaxy as is seen in simulations of CCDM~\cite{Navarro:1996gj}, but rather exhibit a flat core~\cite{Moore:1994yx,Flores:1994gz}. In addition there is the so-called ``missing satellite" problem which is the fact that numerical simulations predict many more dwarf galaxies than what is currently observed in the Milky Way~\cite{Klypin:1999uc,Moore:1999nt,Kauffmann:1993gv}. Furthermore there is the ``too big to fail" problem~\cite{BoylanKolchin:2011de}: it seems that numerical simulations predict dense dwarf galaxies which cannot host the brightest known dwarf galaxies. Although one can claim that the the ``missing satellite" problem may be due to the Milky Way being a statistical fluctuation~\cite{Liu:2010tn,Tollerud:2011wt,Strigari:2011ps}, the ``too big to fail" problem due to unobserved dim dwarf galaxies and the cusp/core problem due to baryonic-DM interactions~\cite{Oh:2010mc,Brook:2011nz,Pontzen:2011ty,Governato:2012fa}, it is possible that the explanation of all the above problems is the existence of sizeable DM-DM interactions.

{The idea that DM self-interactions may ameliorate the aforementioned problems has been studied extensively and in various contexts~\cite{Spergel:1999mh,Wandelt:2000ad,Faraggi:2000pv,Mohapatra:2001sx,Kusenko:2001vu,Loeb:2010gj,Kouvaris:2011gb,Rocha:2012jg,Peter:2012jh,Vogelsberger:2012sa,Zavala:2012us,Tulin:2013teo,Kaplinghat:2013xca,Kaplinghat:2013yxa,Cline:2013pca,Cline:2013zca,Petraki:2014uza,Buckley:2014hja,Boddy:2014yra,Schutz:2014nka}.
In particular, it was pointed out in~\cite{Loeb:2010gj} that DM interacting with a light force carrier $\phi$ that satisfies roughly $\left(m_{X}/10~{\rm GeV}\right) \left(m_{\phi}/100~{\rm MeV}\right)^2 \sim1$ (where $m_{\phi}$ is the mass of the particle $\phi$), can facilitate nicely the flat profile at the core of dwarf galaxies for a range of Yukawa strengths $10^{-5}<\alpha_{X}<1$, while evading constraints on self-interactions coming from galactic and cluster scales.} Although such types of DM self-interactions can resolve some of the problems associated with CCDM, one must ensure that such self-interactions are not strong enough to destroy the ellipticity of spiral galaxies or dissociate the subcluster of the bullet cluster~\cite{Markevitch:2003at}. Depending on the scenario there can be additional strict constraints. For example if $\phi$ couples to the Standard Model through a Higgs portal~\cite{Burgess:2000yq,Patt:2006fw,Andreas:2008xy,Andreas:2010dz,Djouadi:2011aa,Pospelov:2011yp,Greljo:2013wja,Bhattacherjee:2013jca}, one should make sure that $\phi$ decays before the start of the Big Bang Nucleosynthesis (BBN) ($\sim 1$ sec) so the decay products will not affect BBN. This constraint sets a minimum interaction coupling between the Standard Model and the dark sector in order to facilitate the fast decay of $\phi$ before the BBN era. However as was argued in~\cite{Kaplinghat:2013yxa}, the requirement for such a minimum coupling might be at odds with  invisible Higgs decay constraints imposed at LHC and with constraints from direct search experiments like LUX~\cite{Akerib:2013tjd}, since this minimum coupling will lead to a minimum DM-nucleon cross section in underground detectors. In this paper here we show how these problems can be avoided. We will demonstrate that by coupling $\phi$ to neutrinos, we can still have fast decays of $\phi$ in the early universe without leading to violation of the experimental constraints from invisible Higgs decays or direct DM searches. We also provide and study a renormalizable theory where such a model can be realized. 

The paper is organized as follows:  In section \ref{portalpheno} we present the Higgs portal and the relevant constraints that can affect the coupling between the dark sector and the Standard Model. In section \ref{nucouplings} we show how the constraints can be evaded if we couple our mediator to light sterile neutrinos, or to active neutrinos. 
Finally we conclude in section \ref{checkout}.


\section{Higgs Portal Phenomenology}
\label{portalpheno}

\subsection{The Higgs Portal Model}

Consider a scenario in which the dark matter $X$ interacts with a light scalar such that the DM self-scattering cross section is large on dwarf scales, yet small on cluster scales. In general, this scalar $\phi$ will interact with the SM Higgs through the scalar potential. In the case of a real singlet scalar, we consider the {effective} Lagrangian:
\be \mathcal{L} \supset y_{X} \phi \overline{X}X + a \phi |H|^{2} + b \phi^{2} |H|^{2}. 
\label{interactions}
\ee
where $X$ is a Dirac fermion acting as DM, and $H$ is the SM Higgs doublet.
While our essential results depend only on the effective interactions in (\ref{interactions}), it is 
useful to have a simple model realization of these couplings.
A simple and concrete renormalizable  Lagrangian, which leads to the relevant terms of Eq. (\ref{interactions}), is
\be
{\cal L}={\cal L}_{\rm{SM}}+{\cal L}_{\rm{DM}},
\ee
where ${\cal L}_{\rm{SM}}$ is the Standard Model Lagrangian, ${\cal L}_{\rm{DM}}$ is the Lagrangian for a SM singlet real scalar field $S$  and for the dark matter candidate, a SM singlet Dirac fermion, $X$, and their renormalizable interactions.  Explicitly, see e.g.~\cite{Alanne:2014bra},
\beq
{\cal L}_{\rm{DM}} &=& \frac{1}{2}|\partial_\mu S|^2+ \overline{X}i\gamma^\mu\partial_\mu X+y_X S\overline{X}X-V(S,\Sigma),\nonumber \\
V(S,\Sigma) &=& m^2 |\Sigma |^2+\frac{1}{2}m_S^2 S^2+\lambda |\Sigma |^4+\frac{1}{4}\lambda_S S^4 \nonumber \\
&+&\frac{\lambda_{S\Sigma}}{2}|\Sigma |^2S^2+\frac{\mu_3}{3}S^3+\mu_1 |\Sigma |^2 S. \nonumber \\
\label{model}
\eeq
{Where we have assumed that the DM mass arises solely from the vacuum expectation value (VEV) of $S$. There may exist other sources contributing to the mass of $X$, and these can be taken into account by adding to the fermion Lagrangian in Eq.~(\ref{model}) a bare mass term, $m_{X,bare}\bar XX$.}
  The Higgs field $\Sigma$ is written in terms of the electroweak VEV, $v_{EW}$, and fluctuations as
\be
\Sigma=\begin{pmatrix} \sigma^+\\ \frac{1}{\sqrt{2}}(\sigma^0+i\eta^0+v_{\rm{EW}})\end{pmatrix},
\ee
and the singlet as $S=s+w$. Note that we have included the kinetic term of the Higgs field, the usual Higgs potential and the Yukawa interactions with the SM matter fields in ${\cal L}_{\rm{SM}}$.

There are some basic constraints that the parameters in the potential have to satisfy. First,the potential has to be positive at large field values, 
implying that $\lambda>0$ and $\lambda_S>0$. The constraint for $\lambda_{S\Sigma}$ turns out to 
be nontrivial,
$\lambda_{S\Sigma}>-2\sqrt{\lambda\lambda_S}$,
which implies that the coupling $\lambda_{S\Sigma}$ can be negative. 
We are assuming perturbativity in all couplings. Due to the linear interaction $\sim S |\Sigma|^2$ it is inevitable that $w$ must be nonzero whenever $v_{\rm{EW}}$ is. Of course, we are only interested in the case where the global minimum along the neutral Higgs and singlet directions is at $(v_{\rm{EW}},w)$, and we assume that the parameters of the potential are such that this is true. It is convenient to use the minimization condition along $S$ and neutral component of the Higgs field to trade the parameters $m^2$ and $m_S^2$ with the vacuum expectation values $v_{\rm{EW}}$ and $w$. Furthermore, we assume that $w\ll v_{\rm{EW}}$ and that the dimensionful portal coupling $\mu_1\ll v_{\rm{EW}}$.

Under these assumptions, we expand the potential in terms of the vacuum expectation values of the fields and fluctuations around the global minimum $(v_{\rm{EW}},w)$.  The states appearing in Eq. (\ref{interactions}) are related to the fields in Eq. (\ref{model})
\beq
\sigma^0 &=& \cos \theta ~ h+\sin\theta ~\phi,\nonumber \\
s &=& \cos \theta~ \phi-\sin\theta~ h,
\ee
and the physical masses are

\beq
m_\phi^2 &=& \lambda_S w^2+ w\mu_3-\frac{\mu_1}{2w}v_{\rm{EW}}^2,\nonumber \\
m_h^2 &=& 2 \lambda v_{\rm{EW}}^2+\lambda_S w^2  \nonumber \\
&=&2\lambda v_{\rm{EW}}^2+m_\phi^2- w\mu_{3} + \frac{\mu_{1}}{2w} v_{EW}^{2}.
\label{masses}
\eeq

The mixing angle is
\be
\sin\theta\simeq \frac{v_{\rm{EW}}\mu_1}{m_h^2}+\lambda_{S\Sigma}\frac{v_{\rm{EW}}w}{m_h^2},
\label{mixings}
\ee

Finally, we find for the dimension three and four Higgs portal couplings, i.e. the parameters $a$ and $b$ of Eq. (\ref{interactions}), 
\be
a=\mu_1+\lambda_{S\Sigma} w+(6 v_{\rm{EW}}\lambda -2 v_{\rm{EW}}\lambda_{S\Sigma})\sin\theta.
\ee
and $b=\lambda_{S\Sigma}$. Furthermore, due to mixing, the Yukawa coupling between $S$ and the dark matter fermion $X$ now induces a small coupling between the Higgs and the fermion, $\delta_X h \overline{X}X$, where $\delta_X = -y_X\sin\theta$.

{It is natural for the mass scales in the dark sector to be comparable, $\mu_{1} \sim \mu_{3} \sim w$ and much smaller than the EW scale. Adopting this assumption, Eqs.~(\ref{masses}) allow us to express the Higgs self coupling in terms of the physical masses:}
\beq
\lambda&=&\frac{m_h^2}{2v_{\rm{EW}}^2}-\frac{m_\phi^2}{2v_{\rm{EW}}^2}-\frac{\mu_1}{4w} \nonumber \\
&=&\lambda_{\rm{SM}}-\left(\frac{m_\phi^2}{2v_{\rm{EW}}^2}+\frac{\mu_1 y_{X} }{4 \left(m_{X}-m_{X,bare}\right) \cos \theta }\right)  \\ \nonumber
\label{lambdabound}
\eeq

{In the last line we allow for the possibility that the mass of $X$ does not derive entirely from the VEV of $\phi$ but may also include a bare mass contribution.}

Finally imposing vacuum stability, $\lambda >0$, we find that the trilinear coupling $\mu_{1}$ must satisfy
\be 
\mu_{1} < \frac{4 \lambda_{SM} \cos \theta m_{X}}{y_{X}}.
\label{eq:vac}
\ee
We shall return to Eq.~(\ref{eq:vac}) after having examined the constraints imposed by invisible Higgs limits.

%

\subsection{Invisible Higgs Decay}

The Higgs portal Lagrangian in Eq.~(\ref{interactions}) allows for the Higgs to decay invisibly. This leads to two contributions to the invisible width of the Higgs $\Gamma(h\rightarrow {\rm inv}) = \Gamma(h\rightarrow \phi \phi) + \Gamma(h \rightarrow \overline{X}X)$. These contributions are
\bea \Gamma(h\rightarrow \phi \phi) &= &\frac{b^{2}v_{EW}^{2}}{8 \pi m_{h}}\left(1-\frac{4m_{\phi}^2}{m_{h}^{2}}\right)^{1/2}, \\
 \Gamma(h\rightarrow \overline{X}X) &= &\frac{y_{X}^{2}\sin^{2} \theta m_{h}}{8 \pi} \left(1-\frac{4m_{X}^2}{m_{h}^{2}}\right)^{3/2}. \label{higgsdecay}
\eea
Since we will be interested in mediators with mass $m_{\phi} \ll m_{h}$, the process $h\rightarrow \phi \phi$ will always be permitted, though $h\rightarrow \overline{X}X$ is kinematically available only to DM masses $\lesssim 62.5$ GeV. 

To be consistent with the constraints from the LHC on $h\rightarrow {\rm inv}$ we require that the branching be less than $26\%$ (this is the $95\%$ CL limit from which allows for non-SM values of $h \rightarrow \gamma \gamma$ and $gg \leftrightarrow h$ ~\cite{Giardino:2013bma}, though for direct constraints see~\cite{Zhou:2014dba}). Taking the SM contribution to the Higgs width to be $\Gamma_{h,SM}= 4.1$ MeV, we find that Higgs decay into a pair of mediators constrains $b \lesssim 7.1 \times 10^{-3}$, when $m_{\phi} \ll m_{h}$.

In addition the Higgs can also decay directly into a pair of DM particles for $m_{X} < m_{h}/2 $. 
Employing Eqs.~(\ref{mixings}) and (\ref{higgsdecay}) we find 
\be
y_{X} ~{\rm max}\left(a, w\lambda_{S\Sigma} \right) \lesssim \frac{0.9~{\rm GeV}}{\left(1 - \frac{4m_{X}^{2}}{m_{h}^{2}}\right)^{3/4}}
\label{Yukawabound}
\ee
 Thus we see that when $\mu_{1} \simeq a$, the constraint from $h\rightarrow \overline{X}X$ in Eq.~(\ref{Yukawabound}) can be compared against the vacuum stability constraint in Eq.~(\ref{lambdabound}). From this comparison we see that Higgs decay are more stringent than vacuum stability for DM masses $\gtrsim 2$ GeV. If independent bounds for the Higgs self coupling could be derived from e.g. LHC data, this would also provide interesting further constraint on our model.

%
%

\subsection{Thermal Relic Abundance of DM}

Significant DM self-scattering implies that $\phi$ must be light compared to $X$. This gives an automatic contribution to the annihilation cross section of $X$.  {{The particle $\phi$ is kept in equilibrium with the SM bath by processes like $\phi \phi \leftrightarrow h^{*} \leftrightarrow \bar{f}f$ where $f$ is a SM fermion, while $\phi \phi \leftrightarrow \bar{X}X$ keeps DM in equilibrium with $\phi$. {We find that with the coupling $b$ at the upper bound imposed by the Higgs invisible width, the process $\phi \phi \leftrightarrow \mu^{+}\mu^{-}$ is sufficient to keep $\phi$ in equilibrium until DM freeze-out for $m_{X}\gtrsim 1$ GeV.~\footnote{Whether the dark matter sector enters thermal equilibrium in the very early universe in this type of weakly coupled portal models is also nontrivial, see \cite{Enqvist:2014zqa}. }}}  

Thus the annihilation, $\overline{X} X \rightarrow \phi \phi$ will play the dominant role in fixing the relic abundance of DM.  This annihilation process is $p$-wave suppressed
\be \langle \sigma_{\overline{X}X \rightarrow \phi\phi} v_{rel} \rangle = \frac{3y_{X}^{4}}{128 \pi m_{X}^{2}}v^{2} + \mathcal{O}(v^{4}),
\label{thermal}
\ee
implying a strong suppression of indirect signals. The correct thermal relic abundance for a Dirac fermion is obtained for $\langle \sigma v_{rel} \rangle \simeq 4.5\times 10^{-26}~{\rm cm}^{3}{\rm s}^{-1}$~\cite{Steigman:2012nb}. Under the assumption that the two-$\phi$ channel completely dominates, the correct relic abundance is achieved for $y_{X} \simeq 0.43~ \sqrt{m_{X}/100~ {\rm GeV}}$.~\footnote{At large DM masses, the annihilation cross section is Sommerfeld-enhanced, lowering the required value of the coupling for the relic abundance~\cite{Hisano:2004ds,ArkaniHamed:2008qn,Feng:2010zp,Tulin:2013teo}. We take this into account in our figures.} More generally, for models in which the DM carries a particle/antiparticle asymmetry, the annihilation cross section is required to be larger than what is required for the symmetric case. The requisite annihilation cross section in this case depends on the asymmetry and has been calculated in~\cite{Graesser:2011wi} (see also~\cite{Scherrer:1985zt,Griest:1986yu}). 
\begin{figure}[t!]
  \centering
                 \includegraphics[width=0.5\textwidth]{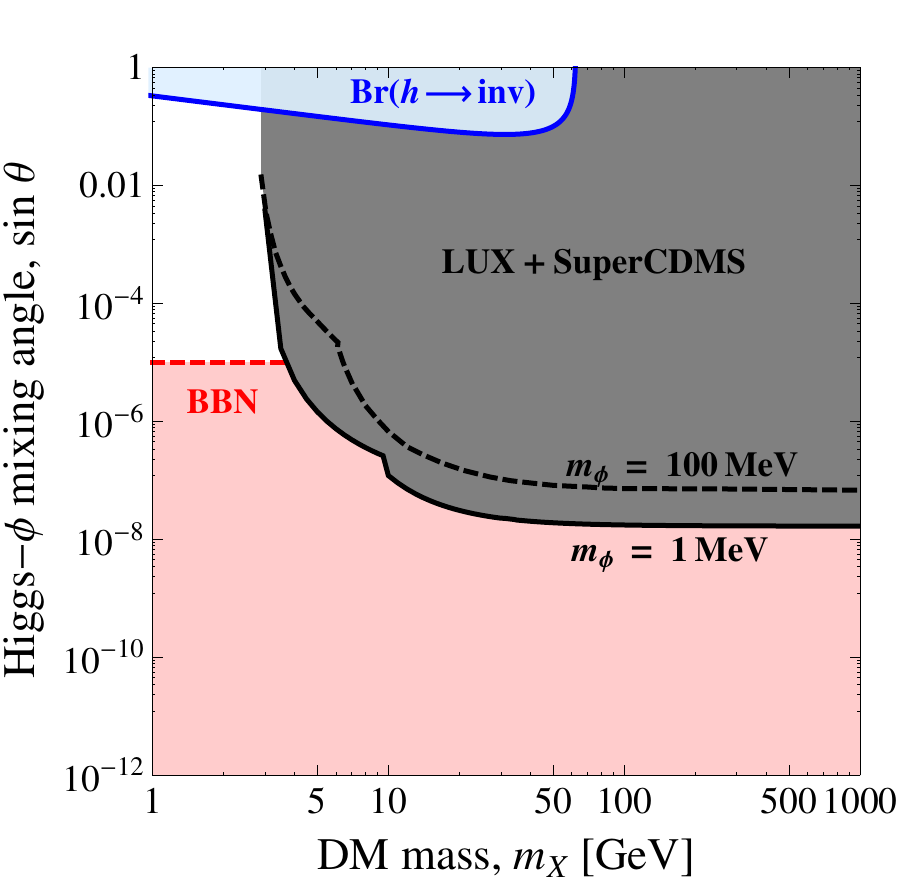}

   \caption{Here we display the LUX~\cite{Akerib:2013tjd} and SuperCDMS~\cite{Agnese:2014aze} constraints on the mixing angle of the Higgs with the SIDM mediator $\phi$ in the minimal model. In addition, we include the constraints from the invisible width of the Higgs and the BBN requirement that $\phi$ decay sufficiently early.
   }
  \label{fig:relic}
\end{figure}

\subsection{Direct Detection Constraints}

{The mass mixing of the singlet scalar $\phi$ with the SM Higgs induces $t$-channel scattering of nuclei, which can be probed at direct detection.  In the Higgs-portal model, the scattering is isospin-conserving and the differential scattering cross section is
\be 
\frac{d\sigma_{XN}}{dE_{R}} \simeq \frac{y_{X}^{2}  m_{N}f_{N}^{2} A^{2}}{\left(m_{\phi}^{2}+q^{2}\right)^{2}} \frac{m_{p}^{2}}{v_{EW}^{2}}\frac{\sin^{2}\theta}{2\pi v^{2}}
\label{eq:dd}
\ee
{where we have assumed $\cos \theta \simeq 1$ and $m_{\phi}\ll m_{h}$, and adopted $f_{N}  \simeq 0.345$~\cite{Cline:2013gha}.}


We can combine the thermal relic requirement Eq.(\ref{thermal}) with the direct detection constraints from LUX~\cite{Akerib:2013tjd} and the recent SuperCDMS~\cite{Agnese:2014aze} data to constrain the $\phi$-$h$ mixing angle $\sin \theta$ for a given choice of $m_{\phi}$. This constraint is illustrated in Fig.~\ref{fig:relic}, where we have chosen two representative values of the mediator mass $m_{\phi} = 1$ and 100 MeV. From the standpoint of direct detection, dropping the mediator mass below 1 MeV does not appreciably impact the limit in Fig.~\ref{fig:relic} because a mediator of this mass is effectively massless for practical purposes at LUX. Increasing the mediator mass beyond 100 MeV on the other hand, has the effect of reducing the direct detection constraints, while simultaneously suppressing the SIDM cross section below the values of interest for dwarf galaxies, see e.g.~\cite{Tulin:2013teo}. 

{However, let us return to the momentum dependence introduced to the cross section when the mediator is light compared the momentum transfer, $m_{\phi} \lesssim q = \sqrt{2m_{N}E_{R}}$.  It is known that this introduces novel features in the recoil spectrum~\cite{Chang:2009yt,Fornengo:2011sz,McDermott:2011hx}. In addition, it has been recently shown these feature cannot be mimicked by DM astrophysics~\cite{Cherry:2014wia}.  Moreover, in~\cite{Cherry:2014wia} it was proposed that future ton-scale experiments can provide an ``astrophysics-free'' determination of DM momentum-dependence include a strong upper bound on the mediator mass.}

\subsection{Tension with BBN Constraints in the Minimal Model}

In contrast with the aforementioned constraints where only upper limits are derived, the cosmology of the mediator $\phi$ requires that the mixing angle be ${\it greater}$ than a critical value. {{Under the assumption that the $\phi$ population is not vastly colder than the visible sector, a substantial number density of $\phi$'s is thermally produced.} To avoid over closure, this therefore requires that $\phi$ be unstable and decay. 

Since self-interactions are most interesting for mediator masses in the 1-100 MeV range, the dominant decay is through the Higgs portal to $\gamma \gamma$ and $e^{+}e^{-}$. However the injection of sizeable energy densities into the visible sector thermal bath after the production of the light elements can spoil the success of BBN. Although the specific BBN constraint depends on the dark sector temperature and the mediator mass, the authors of~\cite{Kaplinghat:2013yxa} argue that simply requiring a $\phi$ lifetime $\lesssim 1$ s is sufficient to evade strong BBN constraints.  

{The decay of $\phi$ proceeds into SM fermions with the rate, $\Gamma_{\phi} = \frac{\sin^{2}\theta y_{f}^{2}m_{\phi}}{8\pi}$, where $y_{f}$ is the Yukawa coupling of the SM fermion $f$. Significant self-scattering at dwarf scales requires mediator masses $\lesssim$100 MeV~\cite{Tulin:2013teo}, implying that the mediator decays dominantly to electrons, $\phi \rightarrow e^{+}e^{-}$. Thus requiring $\Gamma_{\phi}^{-1} < 1$ s in order to satisfy BBN constraints, implies $\sin \theta \gtrsim 10^{-5}$. This lower bound leads to significant tension with the constraints from direct detection. One can see from Fig.~\ref{fig:relic} that this solution to the BBN problem require that the mass of DM is $m_{X} \lesssim 4$ GeV. Future low-threshold direct detection searches and improvements on the invisible Higgs width will further constrain this window at low DM masses.}

{In the following section we explore an alternative solution to the BBN problem that allows for a much wider range of DM masses.}


\section{Coupling $\phi$ to Neutrinos}
\label{nucouplings}
A simple way for $\phi$ to decay sufficiently early in order to satisfy BBN constraints is to introduce a new coupling to neutrinos. This can proceed in one of two ways: (1) $\phi$ decays to a sufficiently light right-handed sterile neutrino via $\phi N^{c}N$, or (2) $\phi$ decays directly to SM neutrinos through the higher dimensional operator $\phi \overline{(LH)} LH$. This latter possibility is natural in models where $\phi$ couples to heavy sterile neutrinos which are then integrated out. Let us examine each possibility in turn. 

\subsection{ Light Sterile Neutrinos} 

The former case is especially interesting given the present hints for $\sim$ eV sterile neutrinos~\cite{Aguilar:2001ty,Aguilar-Arevalo:2013pmq,Mueller:2011nm}. Now with a real scalar $\phi$ interacting with sterile neutrino $N$ through $y_{N} \phi N^{c}N$, we find that for $\phi$ to decay before BBN ($\tau_{\phi} < 1$s), we require
\be y_{N} \gtrsim 6\times 10^{-12} \, \left(\frac{100~{\rm MeV}}{m_{\phi}}\right)^{1/2},
\label{eq:light}
\ee
using $\Gamma_{\phi} = \frac{y_{N}^{2}m_{\phi}}{8 \pi}$. 


Moreover an interaction strength $y_{N} \sim \mathcal{O}(0.1)$ has been advocated\footnote{Although ref.\cite{Dasgupta:2013zpn} considered vector interactions, we expect the contribution from a scalar to be qualitatitively similar.} as a way of suppressing the standard active-to-sterile oscillation production process, easing the cosmological constraints from $N_{{\rm eff}}$~\cite{Dasgupta:2013zpn}. In addition, the authors of~\cite{Aarssen:2012fx,Shoemaker:2013tda,Bringmann:2013vra} have argued that sizeable DM-neutrino couplings could help relieve some of the small-scale structure problems of DM. Though we note that scalar mediators cannot be used to induce late kinetic decoupling since the DM-neutrino scattering cross section scales as $m_{\nu}^{2}$. The lifetime of eV-scale steriles with $\mathcal{O}(1)$ mixing angles is vastly larger than the age of the Universe. Thus together with the well-satisfied $N_{{\rm eff}}$ constraints on their abundance, we conclude that coupling $\phi$ to eV-scale neutrinos provides a satisfactory solution to the otherwise problematic decay of $\phi$. 



For heavier sterile masses, we minimally need the sterile mass to be less than half the $\phi$ mass. 
We must verify that these steriles decay prior to BBN. The decay to 3 active neutrinos dominates over the entire mass range of interest with a rate $\Gamma_{3 \nu} \simeq \frac{G_{F}^{2} m_{N}^{5}}{192 \pi^{3}} \sin^{2} \Theta$, {where $\sin \Theta$ is the sterile-active mixing angle.} For this decay to proceed at pre-BBN times we need 
\be \sin \Theta \gtrsim 8.5 \times 10^{-4} \left(\frac{100~{\rm MeV}}{m_{N}}\right)^{5/2},
\ee
which satisfies the experimental constraints on sterile neutrinos with 3-100 MeV masses~\cite{oai:arXiv.org:0906.2968}.




\subsection{Active Neutrinos}

Next, let us consider the coupling of $\phi$ to the active neutrinos. A simple way the coupling could arise is from $\phi$ coupling to some heavy sterile neutrinos which are then integrated out. If these steriles neutrinos are not too heavy, they can be probed at colliders through the unique decays they produce~\cite{Graesser:2007yj,Graesser:2007pc,Shoemaker:2010fg}.  At low-energy scales the effects of the heavy sterile are encoded in the effective operator, $\phi \left( L H \right)^{2}/\Lambda^{2}$. Then in this case, if we want $\phi$ to decay prior to BBN ($\tau_{\phi} < $ 1s), we need
\be \Lambda \lesssim 2 \times 10^{11}~{\rm GeV} \sqrt{\frac{m_{\phi}}{100~{\rm MeV}}}.
\ee

Such a low scale is may appear somewhat undesirable from the point of view of most UV completions which would also generate the Weinberg operator $y \overline{(LH)} LH/\Lambda$, which upon EW symmetry breaking endows neutrinos with mass, $y v_{EW}^{2}/\Lambda \sim y \times 100$ eV. Thus we require a rather small value for $y \sim 10^{-4}$ in order to reproduce the mass splittings for solar/atmospheric data. {This small value of $y$ may be suggestive of a symmetry reason protecting it. Such models are in contrast with seesaw type models, where instead of tying the smallness of the neutrino masses to large Majorana masses, neutrino masses are naturally small by linking their masses to the small breaking of global lepton number~\cite{Chikashige:1980qk,Gelmini:1980re,Georgi:1981pg,Chacko:2003dt}. } 



In conclusion, we have seen that the BBN constraints can be successfully evaded by simply coupling $\phi$ to neutrinos for a wide range of neutrino masses and properties. This divorces the BBN constraints from the Higgs-$\phi$ mixing allowing us to consider a more general parameter space than~\cite{Kaplinghat:2013yxa}.  Notice that coupling $\phi$ to neutrinos opens another new contribution to the invisible width of the Higgs. However we find that both $y_{N}$ and $\sin \theta$ need to be $\mathcal{O}(1)$ in order for this constraint to become relevant.

\section{Outlook and Conclusions}

{Here we have explored a simple way of reducing the tension between successful BBN and direct detection in Higgs portal DM by introducing a new decay channel for $\phi$ so that it may decay before BBN. Another alternative to reducing the tension between BBN and direct detection is to allow for the possibility of a small Majorana mass for the DM, $Y \langle \phi  \rangle (XX+\bar{X}\bar{X})$ (see e.g.~\cite{Finkbeiner:2007kk,Finkbeiner:2008qu}). Then with $Y \langle \phi \rangle = \mathcal{O}(100~{\rm keV})$ direct detection constraints can be considerably relaxed or removed altogether when the mass splitting is so large that no DM in the halo has sufficient kinetic energy to inelastically scatter.  }
\begin{figure}[t!]
  \centering
                 \includegraphics[width=0.5\textwidth]{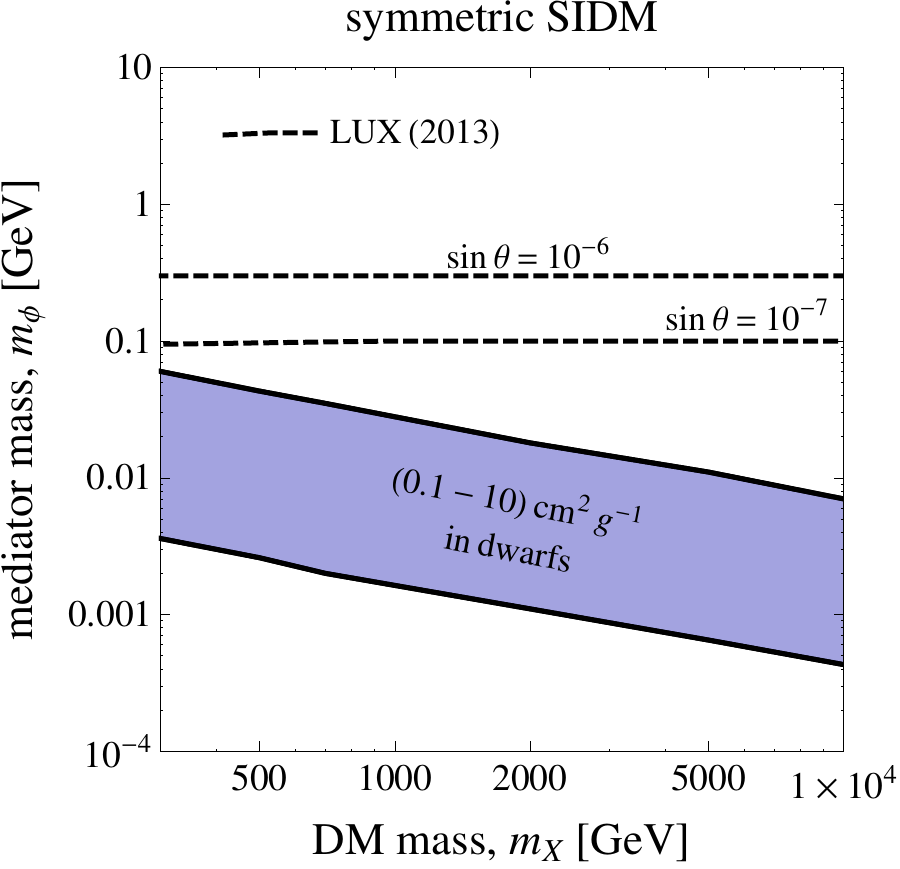}
   \caption{{Here we illustrate the approximate region of interest for addressing the small-scale structure problems of dwarf galaxies, using the results of~\cite{Tulin:2013teo}. The Yukawa coupling $y_{X}$ is fixed by the symmetric thermal relic requirement. The LUX constraints for a fixed choice of the $\phi$-Higgs mixing angle exclude the parameter space {\it below} the dashed black curves. As can be seen from Fig.~\ref{fig:relic} however, sub-$10^{-8}$ mixing angles are not yet constrained by direct detection.}}
  \label{fig:heavySIDM}
\end{figure}

Light mediators can also be probed at low-energy, high-luminosity experiments such as Belle and BaBar~\cite{Batell:2009jf,Schmidt-Hoberg:2013hba,Clarke:2013aya,Essig:2013vha}. Note however that the BaBar mono-photon searches~\cite{Essig:2013vha} are not constraining here since the production of $\phi$ at an $e^{+}e^{-}$ collider is strongly suppressed by the electron Yukawa coupling. Moreover, constraints from the invisible width of the upsilon, which require $\mathcal{BR}\left(\Upsilon(1{\rm S})\rightarrow ``{\rm invisible}"\right) < 3 \times 10^{-4}$, impose the mild constraint $y_{N}^{2} \sin^{2} \theta  \lesssim 0.81$ (assuming $m_{\phi} \ll m_{\Upsilon}$). 

{Outside of the resonant regime near $\sim$ 10-100 MeV~\cite{Kaplinghat:2013yxa}, the parameter space relevant for SIDM can be broken into two qualitatively distinct regimes: one at heavy DM mass where self-scattering proceeds mostly in the classical non-perturbative regime, and and a light DM regime where the scattering proceeds mostly in the perturbative Born regime~\cite{Tulin:2013teo}. In the light DM regime, future probes include invisible and visible Higgs decays proceeding from $h\rightarrow \bar{X}X$ and $h\rightarrow \phi \phi$. Portions of the parameter space may be probed additionally at direct detection, $m_{X} \gtrsim 4$ GeV.}

{In the heavy DM regime, the requirement of SIDM imposes the need for rather light mediators. {For this to be consistent with direct detection constraints, we require small mixing angles, $\sin \theta \lesssim 10^{-8}$.  This is illustrated in Fig.~\ref{fig:heavySIDM}.  However, notice that we want very small mixing angles, but also need the parameter $b$ in Eq.~(\ref{interactions}) to be $ \gtrsim 10^{-3}$ in order for $\phi$ to stay in thermal equilibrium with the bath through the freeze-out of $X$.  We see from Eq.(\ref{mixings}) that we then need the fine-tuning, $\mu_{1} \simeq -bw$. This allows for both small mixing angles while taking the parameter $b$ relatively large. This tuning becoming much less severe going to lower DM masses, where the mixing angle is allowed to be larger (see e.g. Fig~\ref{fig:relic}). }

However the large DM and low mediator masses in this regime offer the novel possibility of detecting a signal of the presence of a light mediator~\cite{Cherry:2014wia}.  When the mediator mass $m_{\phi}$ is light compared to the momentum transfer $q$, the non-trivial momentum dependence of the scattering in direct detection can lead to distinctive recoil spectra~\cite{Chang:2009yt,Fornengo:2011sz,McDermott:2011hx}.  If two or more direct detection experiments with disparate targets observe a signal, the combined data may yield an astrophysics-free indication of the existence of a light mediator~\cite{Cherry:2014wia}. Regions of the parameter space with this sensitivity will have low mixing angle, $\sin \theta < 10^{-7}.$ }


\label{checkout}

We have seen that phenomenology of self-interacting DM through the Higgs portal is quite rich. In this model a new singlet scalar provides DM with a velocity-dependent self-scattering cross section, which can well-satisfy the desire to have large self-interactions at dwarf scales while satisfying the constraints from cluster scales. In the minimal setup where the scalar only interacts with DM and the SM Higgs there is considerable tension with the combination of BBN and direct detection constraints. We focused on a simple solution to evading the BBN bounds, which is to endow the singlet scalar with interactions with sterile neutrinos. This has the welcome benefit of allowing the scalar to decay well-before the BBN epoch, thereby divorcing BBN constraints from the parameters relevant for direct detection.  We stress that future probes of the model will come from both low-threshold and larger-exposure direct detection, improved limits on the invisible decay width of the Higgs. More model-dependent signatures include absorption features in the high-energy neutrino spectrum at IceCube and other probes of neutrino self-interactions~\cite{Cyr-Racine:2013jua,Cherry:2014xra}.
\\

\acknowledgements

This work has been supported by the CP3-Origins centre which is partially funded by the Danish National Research Foundation, grant number DNRF90.

\bibliographystyle{ArXiv}
\bibliography{DM.bib}

\end{document}